\documentclass[fleqn,10pt]{article}

\usepackage{amsmath,amssymb}
\usepackage{graphicx,verbatim}

\title{Precision spectroscopy of the hydrogen molecular ions: present status of theory and experiment}
\author{V.I.Korobov\\[2mm]
Bogolyubov Laboratory of Theoretical Physics, \\ Joint Institute for Nuclear Research, Dubna, 141980, Russia}

\date{}

\begin{document}

\maketitle

\begin{abstract}
We review recent experiments on HD$^+$ spectroscopy. The reduced proton-deuteron mass to electron mass ratio, $\mu_{pd}/m_e$, is inferred from the comparison of theory and experiment. Theoretical issues related to the calculations of the spin structure, which are currently the main limiting factor of theoretical accuracy, are discussed.
\end{abstract}

\vspace{3mm}
Contribution to the Proceedings of the Conference on Precision Physics and Fundamental Physical Constants, Slovakia, 2021.


\section{Introduction}

In recent years substantial progress in the precision spectroscopy of the HD $^+$ molecular ion through Doppler-free spectroscopy of ultracold trapped ions in the Lamb-Dicke regime \cite{Schiller20,Patra20,Kortunov21} has been achieved.

This conference proceedings contribution has a threefold interest: we want to present an analysis of the obtained experimental data and to infer from it the new value of the reduced mass of the two nuclei: $m_pm_d/(m_p+m_d)$ normalized to the electron mass.

Then we give an independent derivation of the pure rotational spin-averaged transition frequency based on the experimental data of the ($G_1=1,G_2=2$) manifold of spin states. We keep here the notation for the spin coupling similar to Ref.~\cite{Schiller20}: $\mathbf{G}_1=\mathbf{s}_e\!+\!\mathbf{I}_p$, $\mathbf{G}_2\!=\!\mathbf{G}_1\!+\!\mathbf{I}_d$, and $\mathbf{F}\!=\!\mathbf{G}_2\!+\!\mathbf{N}$, where $\mathbf{N}$ is the total orbital angular momentum operator.

Finally we consider the spin interaction HFS Hamiltonian and will show that the contributions from the \emph{nuclear spin -- electron spin -- nuclear spin} interactions of $m\alpha^6$ order are by far lower then the present theoretical and experimental uncertainty.

In our analysis, we will rely on the theoretical results obtained by our group in \cite{Korobov17,Korobov21}.

\section{Nuclear $\mu_{pd}$ reduced mass from the HD$^+$ spectroscopy.}

\begin{table}[t]
\caption{Calculated sensitivity coefficients $A_i$ for Eq.~(\ref{sens}).}\label{tab:A}
\begin{center}
\begin{tabular}{l@{\hspace{15mm}}c}
\hline\hline
transition   & $(2cR_\infty)A_i\!\times\!(\mu/\nu_{0,i})$ \\
\hline
$\nu_{(00)\to(10)}$ & $-$0.99 \\
$\nu_{(00)\to(11)}$ & $-$0.49 \\
$\nu_{(30)\to(39)}$ & $-$0.35 \\
\hline\hline
\end{tabular}
\end{center}
\end{table}

\begin{table}[b]
\caption{Theoretical and experimental spin-averaged transition frequencies (in kHz). CODATA18 values of fundamental constants were used in the calculations.}
\label{Tab1}
\begin{center}
\begin{tabular}{c@{\hspace{3mm}}c@{\hspace{3mm}}c}
\hline\hline
$(N,v)\to(N',v')$ & theory & experiment \\
\hline
$(0,0)\to(1,0)$ &  1\,314\,925\,752.932(19)  & 1\,314\,925\,752.910(17) \\
$(0,0)\to(1,1)$ & 58\,605\,052\,163.9(0.5)  & 58\,605\,052\,164.24(86) \\
$(3,0)\to(3,9)$ & 415\,264\,925\,502.8(3.3) & 415\,264\,925\,501.8(1.3) \\
\hline\hline
\end{tabular}
\end{center}
\end{table}

First, we want to make an independent calculation of the reduced mass of the nuclei in the HD$^+$ molecular ion from the experimental data. In fact, the spectral transitions in the hydrogen molecular ions may be obtained using the adiabatic approximation with precision of about five significant digits, that means that the spectral line frequencies as functions of the fundamental constants depend mainly on the reduced mass and only very few on other configurations of the mass parameters. More precisely, we perform numerical calculations of the nonrelativistic three-body problem using atomic units, and thus the parameter
\begin{equation}
\mu = \frac{\mu_{pd}}{m_e}=\frac{m_pm_d}{(m_p+m_d)m_e}
\end{equation}
and the Rydberg constant which is known with relative precision $1.2\times10^{-12}$ from the last CODATA adjustment \cite{CODATA18} and the only parameters determining the transition frequencies in HD$^+$. The latter may be used in our analysis as an exact constant. The uncertainty in other fundamental constants like $r_p$ and $r_d$ contribute to the relative level much below $10^{-12}$.

In order to get the experimental value of $\mu$ we linearize the dependence of the frequency of the transition line as a function of the fitted parameters:
\begin{equation}\label{sens}
\nu_{{\rm exp},i}(R_\infty,\mu)=\nu_{0,i}+\frac{dR_\infty}{R_\infty}\,\nu_{0,i}+(2cR_\infty)A_i\,d\mu.
\end{equation}
The coefficients $A_i$ are obtained numerically (see Table \ref{tab:A}) from the nonrelativistic Schr\"o\-din\-ger three-body equation using the Hellman-Feynman theorem \cite{Weinberg}. As is seen from the Table the properly normalized coefficients correspond to the molecular spectroscopy picture: a rotational transition is proportional to $1/\mu$ and vibrational transition to $\sqrt{1/\mu}$. Only in the case of the $(3,0)\!\to\!(3,9)$ transition, the coefficient becomes smaller than 0.5 since the upper state is already in that part of the potential well where anharmonicity is important.

Results of recent experiments and corresponding theoretical values are presented in Table \ref{Tab1}. Next Table (Table \ref{tab:red}) shows our results for the reduced mass taken from three individual experimental data and the last line is the statistical average of the three experiments. Relative uncertainty of the obtained value is $u_r(\mu)=1.4\times10^{-11}$.
\begin{table}[h]
\caption{Reduced mass $\mu$ inferred from the HD$^+$ ion spectroscopy}\label{tab:red}
\begin{center}
\begin{tabular}{cc}
\hline
 & $\mu$ \\
\hline
CODATA18        & $1223.899\,228\,722(51)$\hspace*{13mm} \\
\hline
$(0,0)\to(0,1)$ & $1223.899\,228\,743(16)_{\rm exp}(17)_{\rm th}$  \\
$(0,0)\to(1,1)$ & $1223.899\,228\,707(17)_{\rm exp}(17)_{\rm th}$  \\
$(3,0)\to(3,9)$ & $1223.899\,228\,730(04)_{\rm exp}(17)_{\rm th}$  \\
\hline
                & $1223.899\,228\,730(04)_{\rm exp}(17)_{\rm th}$  \\
\hline
\end{tabular}
\end{center}
\end{table}

If we combine this result with the high precision mass spectrometry experiment carried out by Myers' group at Florida State University \cite{Fink20}:
\[
m_d/m_p = 1.999\,007\,501\,274(38),
\]
we arrive at new values for the proton-electron and deuteron-electron mass ratios:
\[
m_p/m_e = 1836.152\,673\,476(44),
\qquad
m_d/m_e = 3670.482\,967\,763(88).
\]

\section{Yet another determination of the spin-averaged frequency for the $(N\!=\!0,v\!=\!0)\to(1,0)$ transition}

In the first experiment at D\"usseldorf \cite{Schiller20} six HFS transitions were measured. Three of them belong to the manifold of the same spin structure and different coupling with orbital angular momentum $N$. Thus the weighted sum of the three states
\begin{equation}
\nu_{\rm spin} = \frac{1}{15}\sum_{F=1}^3(2F+1)f^{\rm exp}(122\to12F)-f^{\rm exp}_{\rm spin-av}
\end{equation}
should not depend on the HFS couplings with $\mathbf{N}$ and may be calculated from theory \cite{Karr_HFS20} using the spin-spin coupling constants $E_4$ and $E_5$ only, which are accurate up to six significant digits. Thus the corresponding theoretical value for this shift is
\[
\nu_{\rm spin}^{\rm theor} = 1971.759(2),
\]
and the spin-averaged frequency for the $(N\!=\!0,v\!=\!0)\to(1,0)$ transition deduced from this analysis is equal to
\begin{equation}
f^{\rm exp}_{\rm spin-av} = 1\,314\,925\,752.892(25)_{\rm exp}(2)_{\rm theor,spin} \mbox{ kHz},
\end{equation}
which is in a good agreement with the previous \emph{composite frequency} result. On the other hand, this trick eliminates the critical inaccuracy in the HFS spin-orbit coupling coefficients.

\section{The nuclear spin-spin interaction}

The nuclear spin-spin interaction,
\begin{equation}
\delta \left( \mathbf{I}_p\!\cdot\!\mathbf{I}_d \right),
\end{equation}
via the second-order electron coupled interactions has been considered by Ramsey in \cite{Ramsey53}. As was discussed in \cite{Ramsey53}, the largest contribution comes from the \emph{electron spin--nuclear spin} interaction. In our case it may be expressed as
\begin{equation}\label{NN_SO}
\Delta E_{s_Ns_N}^{(0)} =
   2\left\langle
      H_{ss;p}^{(0)} Q (E_0-H_0)^{-1} Q H_{ss;d}^{(0)}
   \right\rangle,
\end{equation}
where
\[
H_{ss;a}^{(0)} = -\frac{16\pi}{3}\mu_e\frac{\mu_a}{I_a}\mu_0\mu_N(\mathbf{s}_e\!\cdot\!\mathbf{I}_a).
\]

Then using
\[
(\boldsymbol{\sigma}_e\!\cdot\!\mathbf{I}_p)(\boldsymbol{\sigma}_e\!\cdot\!\mathbf{I}_d)
 = (\mathbf{I}_p\!\cdot\!\mathbf{I}_d)+i\boldsymbol{\sigma}_e[\mathbf{I_p}\!\times\!\mathbf{I_d}]
\]
one may get a new term into the effective HFS Hamiltonian, where the nuclear spin-spin interaction appears explicitly. However, in our case it is more convenient to keep the interaction in the form:
\begin{equation}
H_{N\!N} = \delta_{N\!N}\bigl[(\mathbf{s}_e\mathbf{I}_p)(\mathbf{s}_e\mathbf{I}_d)+(\mathbf{s}_e\mathbf{I}_d)(\mathbf{s}_e\mathbf{I}_p)\bigr]/2.
\end{equation}
Then performing the numerical calculations, which average the spatial part of Eq.~(\ref{NN_SO}), we obtain for the $(N\!=\!1,v\!=\!0)$ state of HD$^+$ ion that the coefficient
\[
\delta_{N\!N} = 117 \mbox{ Hz}.
\]
This leads to a shift in the energy of an individual HFS state to less than 27 Hz. In its turn, for the measured HFS transition energies of the pure rotational transition \cite{Schiller20}, the corresponding correction is less than 2 Hz, which may be neglected at present.

\section{Conclusion}

In conclusion we want to state that the new precision spectroscopy data may be used for determination of the proton-electron and deuteron-electron mass ratio, which provides a three and two-fold improvement with respect to CODATA18 values. Next we have shown that inclusion of the nuclear spin-spin interaction into the HFS Hamiltonian does not solve the problem of some discrepancy between experiment and theory in the hyperfine structure of the HD$^+$ ion.

\section*{Acknowledgements}

This work was done in collaboration with Laurent Hilico, Jean-Philippe Karr, Mohammad Haidar (LKB), and Zhen-Xiang Zhong (Wuhan, WIPM CAS) that is gratefully acknowledged.

\end{document}